\documentstyle[erice98,12pt,graphicx]{article}

\def\Journal#1#2#3#4{{#1} {\bf #2} (#4) #3}

\def\NPA{{\em Nucl. Phys.} A}

\def\PLB{{\em Phys. Lett.} B}

\def\PRL{\em Phys. Rev. Lett.}

\def\PRD{{\em Phys. Rev.} D}
\def\PRC{{\em Phys. Rev.} C}

\def\ZPA{{\em Z. Phys.} A}

\begin{document}

\title{
HADRONS IN THE NUCLEAR MEDIUM\footnote{work supported by BMBF and DFG}}

\author{U. MOSEL\footnote{mosel@theo.physik.uni-giessen.de}}
\address{Institut fuer Theoretische Physik, Universitaet Giessen\\
D-35392 Giessen, Germany}


\maketitle

\abstracts{In this talk I first discuss predictions based on QCD sum
rules and on hadronic models for the properties of vector mesons in the
nuclear medium. I then describe possible experimental signatures and
show detailed predictions for dilepton invariant mass spectra with a
special emphasis on nuclear reactions involving elementary incoming
beams and nuclear targets. I will in particular illustrate that the
sensitivity of pion and photon induced reactions to in-medium effects
is nearly as large as that of heavy-ion reactions.}

\section{Introduction}
The investigation of so-called in-medium properties of hadrons has
found widespread interest during the last decade. This interest was
triggered by two aspects.

The first aspect that triggered an interest in the investigation of
in-medium properties of hadrons was a QCD sum-rule based prediction by
Hatsuda and Lee \cite{HL} in 1992 that the masses of vector mesons
should drop dramatically as a function of nuclear density. It was widely
felt that an
experimental verification of this prediction would establish a
long-sought direct link between quark degrees of freedom and nuclear
hadronic interactions. In the same category fall the predictions of
Brown and Rho that argued for a general scaling for hadron masses with
density \cite{Brown-Rho}.

The second aspect is that even in ultrarelativistic heavy-ion reactions,
searching for
observables of a quark-gluon plasma phase of nuclear matter, inevitably
also many relatively low-energy ($\sqrt{s} \approx 2 - 4 \:
\mbox{GeV}$) final state interactions take place. These interactions
involve, for example, collisions between many mesons for which the
cross-sections and meson self-energies in the nuclear medium are not
known, but may influence the interpretation of the experimental
results.

Hadron properties in medium involve masses, widths and coupling
strengths of these hadrons. In lowest order in the nuclear density all
of these are linked by the $t \rho$ approximation that assumes that the
interaction of a hadron with many nucleons is simply given by the
elementary $t$-matrix of the hadron-nucleon interaction multiplied with
the nuclear density $\rho$. For vector mesons this approximation reads
\begin{equation}\label{Vself}
\Pi_{\rm V} = - 4 \pi f_{\rm VN} (0) \rho
\end{equation}
where $f_{\rm VN}$ is the forward-scattering amplitude of the vector
meson (V) nucleon (N) interaction. Approximation (\ref{Vself}) is good
for low densities ($\Pi_{\rm V}$ is linear in $\rho$) and/or large
relative momenta where the vector meson `sees' only one nucleon at a
time. Relation (\ref{Vself}) also neglects the Fermi-motion of the
nucleons although this could easily be included.

Simple collision theory \cite{EJ,Kon} then gives the shift of mass and
width of a meson in nuclear matter as
\begin{eqnarray}\label{masswidth}
  \delta m_{\rm V} &=& - \gamma v \sigma_{\rm VN} \eta \rho \nonumber \\
  \delta \Gamma_{\rm V} &=& \gamma v \sigma_{\rm VN} \rho ~.
\end{eqnarray}
Here, according to the optical theorem, $\eta$ is given by the ratio of
real to imaginary part of the forward scattering amplitude
\begin{equation}
  \eta = \frac{\Re{f_{\rm VN}(0)}}{\Im{f_{\rm VN}(0)}} ~.
\end{equation}

The expressions (\ref{masswidth}) are interesting since an
experimental observation of these mass- and width-changes could give
valuable information on the free cross sections $\sigma_{\rm VN}$ which
may not be available otherwise. The more fundamental question, however, is if
there is more to in-medium properties than just the simple collisional
broadening predictions of (\ref{masswidth}).

\section{Fundamentals of Dilepton Production}

From QED it is well known that vacuum polarization, i.e. the virtual
excitation of electron-positron pairs, can dress the photon. Because
the quarks are charged, also quark-antiquark loops can dress the photon.
These virtual
quark-antiquark pairs have to carry the quantum numbers of the photon,
i.e.\ $J^\pi = 1^-$. The $q \bar q$ pairs can thus be viewed as vector
mesons which have the same quantum numbers; this is the basis of
Vector Meson Dominance (VMD).

The vacuum polarization tensor is then, in complete analogy to QED,
given by
\begin{equation}\label{jjcorr}
  \Pi^{\mu \nu} = \int d^4x \, e^{i q x}
  \langle 0 | T[j^\mu (x) j^\nu (0)] | 0 \rangle
  = \left( g^{\mu \nu} - \frac{q^\mu q^\nu}{q^2} \right) \Pi(q^2)
\end{equation}
where $T$ is the time ordering operator.
Here, in the second line, the tensor structure has been exhibited
explicitly. This so-called current-current correlator contains the
currents $j^\mu$ with the correct charges of the vector mesons in
question. Simple VMD \cite{Sak} relates these currents to the vector
meson fields
\begin{equation}\label{VMD}
  j^\mu(x) = \frac{{m_{\rm V}^0}^2}{g_{\rm V}} V^\mu(x)~.
\end{equation}
Using this equation one immediately sees that the current-current
correlator (\ref{jjcorr}) is nothing else but the vector meson
propagator $D_{\rm V}$
\begin{equation}\label{Vprop}
  \Pi(q^2) = \left( \frac{{m_{\rm V}^0}^2}{g_{\rm V}} \right)^2
             D_{\rm V} (q^2) ~.
\end{equation}
The scalar part of the vector meson propagator is given by
\begin{equation}\label{Vprop1}
D_{\rm V}(q^2) = \frac{1}{q^2 - {m_{\rm V}^0}^2 - \Pi_{\rm V}(q^2)}~.
\end{equation}
Here $\Pi_{\rm V}$ is the selfenergy of the vector meson.

For the free $\rho$ meson information about $\Pi(q^2)$ can be obtained
from hadron production in $e^+ e^-$ annihilation reactions \cite{PS}
\begin{equation}\label{hadprod}
 R(s) = \frac{\sigma (e^+ e^- \rightarrow \mbox{hadrons})}
             {\sigma(e^+ e- \rightarrow \mu^+ \mu^-)}
      = - \frac{12 \pi}{s} \Im \Pi(s) ~
\end{equation}
with $s = q^2$. This determines the imaginary part of $\Pi$ and,
invoking vector meson dominance, also of $\Pi_{\rm V}$. The data (see,
e.g. Fig.\ 18.8 in \cite{PS}, or Fig.\ 1 in \cite{KW}) clearly show at
small $\sqrt{s}$ the vector meson peaks, followed by a flat plateau
starting at $\sqrt{s} \approx 1.5 \: \mbox{GeV}$
described by perturbative QCD.

In order to get the in-medium properties of the vector mesons, i.e.\
their selfenergy $\Pi_{\rm V}$, we now have two ways to proceed: We
can, \emph{first}, try to determine the current-current correlator by
using QCD sum rules \cite{HL}; from this correlator we can then
determine the self-energy of the vector meson following
eqs.\ (\ref{Vprop}),(\ref{Vprop1}). The \emph{second} approach consists
in setting up a hadronic model and calculating the selfenergy of the
vector meson by simply dressing its propagators with appropriate
hadronic loops. In the following sections I will discuss both of these
approaches.

\subsection{QCD sum rules and in-medium masses}

The QCD sum rule for the current-current correlator is obtained by
evaluating the function $R(s)$, and thus $\Im \Pi(s)$ (see
(\ref{hadprod})), in a hadronic model on one hand and in a QCD-based
model on the other. The latter, QCD based, calculation uses the fact
that the current-current correlator (\ref{jjcorr}) can be Taylor
expanded in the space-time distance $x$ for small space-like distances
between $x$ and $0$; this is nothing else than the Operator Product
Expansion (OPE) \cite{PS}. In this way we obtain for the free meson
\begin{equation}
R^{\rm OPE}(M^2) = \frac{1}{8 \pi^2} \left(1 + \frac{\alpha_{\rm
S}}{\pi} \right) + \frac{1}{M^4} m_{\rm q} \langle \bar{q} q \rangle +
\frac{1}{24 M^4} \langle \frac{\alpha_{\rm S}}{\pi} G^2 \rangle -
\frac{56}{81 M^6} \pi \alpha_{\rm S} \kappa \langle \bar{q} q \rangle^2
~.
\end{equation}
Here $M$ denotes the so-called Borel mass. The expectation values
appearing here are the quark- and gluon-condensates. The last term here
contains the mean field approximation
\begin{equation}
\langle (\bar{q} q)^2 \rangle = \kappa \langle \bar{q} q \rangle^2 ~.
\end{equation}

The other representation of $R$ in the space-like region can be
obtained by analytically continuing $\Im \Pi(s)$ from the time-like to
the space-like region by means of a twice subtracted dispersion
relation. This finally gives
\begin{equation}
R^{\rm HAD} (M^2) = \frac{\Re{\Pi^{\rm HAD}(0)}}{M^2}
 - \frac{1}{\pi M^2} \int_0^\infty
ds \, \Im \Pi^{\rm HAD}(s) \frac{s}{s^2 + \epsilon^2}
\exp{-s/M^2}~.
\end{equation}
Here $\Pi^{\rm HAD}$ represents a phenomenological hadronic spectral
function. Since for the vector mesons this spectral
function is dominated by resonances in the low-energy part it
is usually parametrized in terms of a resonance part
with parameters such as strength, mass and width at low energies with a
connection to the QCD perturbative result for the quark structure for
the current-current correlator at higher energies (for details
see the contribution by S. Leupold et al. \cite{Leupoldbar} in these
proceedings and refs. \cite{Leup1,Leup2,Lee}).

The QCD sum rule is then obtained by setting
\begin{equation}
R^{\rm OPE}(M^2) = R^{\rm HAD}(M^2)~.
\end{equation}
Knowing the lhs of this equation then allows one to determine the
parameters in the spectral function appearing in $R^{\rm HAD}$ on the
rhs. If the vector meson moves in the nuclear medium, then $R$ depends also
on its momentum. However, detailed studies \cite{Leup2,Lee} find only a very
weak momentum dependence.

The first applications \cite{HL} of the QCDSR have used a simplified
spectral function, represented by a $\delta$-function at the meson mass
and a perturbative QCD continuum, starting at about $s \approx 1.2$
GeV. Such an analysis gives a value for the free meson mass that agrees
with experiment. On this basis the QCDSR has been applied to the
prediction of in-medium masses of vector mesons by making the
condensates density-dependent (for details see \cite{HL,Leup1,Leup2}).
This then leads to a lowering of the vector meson mass in nuclear
matter.

This analysis has recently been repeated with a spectral function that
uses a Breit-Wigner parametrization with finite width. In this
study \cite{Leup1} it turns out that QCD sum rules are compatible with a
wide range of masses and widths (see also ref. \cite{KW}). Only if the
width is -- artificially -- kept zero, then the mass of the vector meson
has to drop with
nuclear density \cite{HL}. However, also the opposite scenario, i.e. a
significant broadening of the meson at nearly constant pole position,
is compatible with the QCDSR. 

\subsection{Hadronic models}

Hadronic models for the in-medium properties of hadrons start from
known interactions between the hadrons and the nucleons. In principle,
these then allow one to calculate the forward scattering amplitude
$f_{\rm VN}$ for vector meson interactions, for example. Many such
models have been developed over the last few
years \cite{KW,HFN,AK,FP,Rapp}.

The model of Friman and Pirner \cite{FP} was taken up by Peters et
al. \cite{Peters} who also included
$s$-wave nucleon resonances. It turned out that in this analysis the
$D_{13}\: N(1520)$ resonance plays an overwhelming role. This resonance
has a significant $\rho$ decay branch of about 20 \%. Since at the pole
energy of 1520 MeV the $\rho$ decay channel is not yet energetically
open this decay can
only take place through the tails of the mass distributions of
resonance and meson. The relatively large relative decay branch then
translates into a very strong $N^* N \rho$ coupling constant (see
also \cite{PE,Rapp}).

\begin{figure}
  \centerline{\includegraphics[height=6cm]{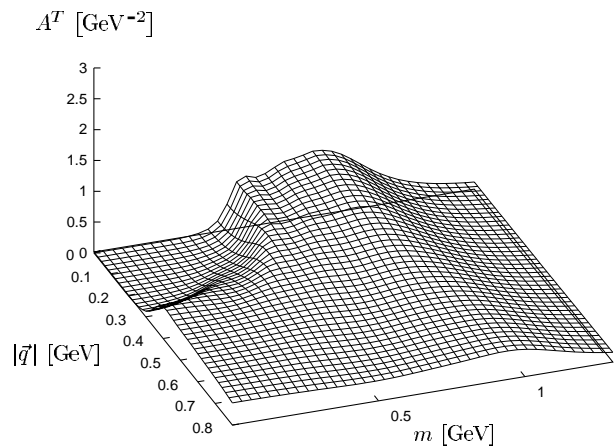}}
  \caption{Spectral function of transverse $\rho$ mesons as a function of
  their three-momentum and invariant mass (from \protect\cite{Peters}).}
  \label{Figrhot}
\end{figure}

\begin{figure}
  \centerline{\includegraphics[height=6cm]{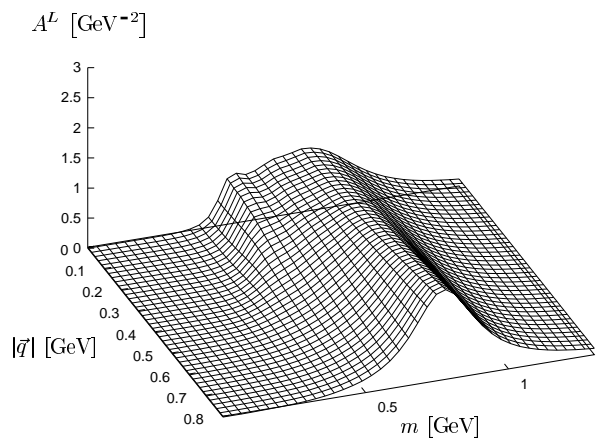}}
  \caption{Spectral function of longitudinal $\rho$ mesons as a function
  of their three-momentum and invariant mass (from \protect\cite{Peters}).}\label{Figrhol}
\end{figure}

The main result of this $N^* h$ model for the $\rho$ spectral
function is a considerable broadening for the latter. This
is primarily so for the transverse vector mesons (see Fig. \ref{Figrhot}),
whereas the longitudinal degree of freedom gets only a little broader with
only a slight change of strength downwards (see Fig.\ \ref{Figrhol}\cite{Peters}.

The results shown in Fig.\ \ref{Figrhot} actually go beyond the simple "$t\rho$"
approximation discussed earlier (see (\ref{Vself})) in that they
contain higher order density effects: a lowering of the $\rho$ meson
strength leads to a strong increase of the phase-space available for
decay of the $N(1520)$ resonance; this causes a corresponding strong
increase of the $N(1520)$ $\rho$-decay width which in turn affects the
spectral function. The result is the very broad, rather featureless spectral function for the
transverse $\rho$ shown in Fig.\ \ref{Figrhot}.

\section{Experimental Observables}
In this section I will now discuss various possibilities to verify
experimentally the predicted changes of the $\rho$ meson properties in
medium.

\subsection{Heavy-Ion Reactions}

Early work \cite{Wolf,XK} on an experimental verification of the
predicted broadening of the $\rho$ meson spectral function has
concentrated on the dilepton spectra measured at relativistic energies
(about 1 -- 4 A GeV) at the BEVALAC, whereas more recently many
analyses have been performed for the CERES and HELIOS data obtained at
ultrarelativistic energies (150 -- 200 A GeV) at the SPS. In such collisions
nuclear densities of about 2 - 3 $\rho_0$ can
already be reached in the relativistic domain; in the ultrarelativistic
energy range baryon densities of up to $10 \rho_0$ are predicted (for a
recent review see \cite{BC}). Since the selfenergies of produced vector
mesons are at least proportional to the density $\rho$ (see
(\ref{Vself})) heavy-ion reactions seem to offer a natural
enhancement factor for any in-medium changes.

\begin{figure}
\centerline{\includegraphics[height=9cm]{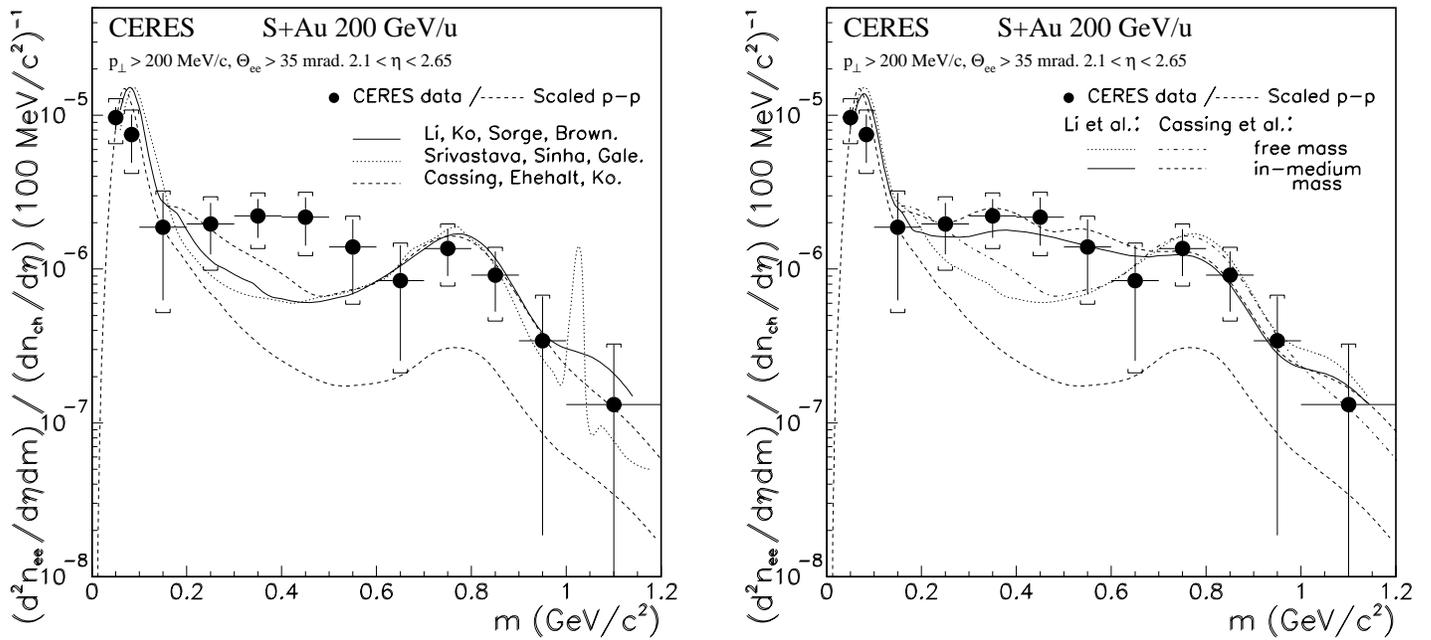}}
\caption{Comparison of several transporttheoretical calculations of the
dilepton invariant mass spectrum with the CERES data. The calculations on
the left show results of calculations employing free hadron properties, those
on the right employ in-medium corrections to the $\rho$ meson (solid and
dashed lines) (from \protect\cite{Heros}).} \label{ceresfig}
\end{figure}

The CERES data \cite{CERES} indeed seem to confirm this expectation
(see Fig.\ \ref{ceresfig}).
The left calculation shows that a large part of the CERES dilepton yield
can already be explained by simple secondary reactions, mainly by $\pi^+\pi^-
\rightarrow e^+e^-$ because of the known high pion multiplicity in such
high-energy heavy-ion reactions. Even then, there is, however, some yield
still missing. The present situation is -- independent of
the special model used for the description of the data -- that agreement
with the measured dilepton mass spectrum in the mass range between
about 300 and 700 MeV for the 200 A GeV $S + Au$ and $S + W$ reactions
can only be obtained if $\rho$-meson strength is shifted downwards (for
a more detailed discussion see \cite{BC,QM}) (for the recently measured 158
A GeV $Pb + Au$ reaction the situation is not so clear; here the
calculations employing `free' hadron properties lie at the lower end of
the experimental error bars \cite{BC}).

However, all the
predictions are based on equilibrium models in which the properties of
a $\rho$ meson embedded in nuclear matter with infinite space-time
extensions are calculated. A ultrarelativistic heavy-ion collision is
certainly far away from this idealized scenario. In addition, heavy-ion
collisions necessarily average over the polarization degrees of freedom.
The two physically
quite different scenarios, broadening the spectral
function or shifting simply the $\rho$ meson mass downwards while keeping
its free width, thus lead to indistinguishable observable consequences
in such collisions.
This can be understood by observing that even in an ultrarelativistic
heavy-ion collision, in which very high baryonic densities are reached,
a large part of the observed dileptons is actually produced at rather
low densities (see Fig. 3 in \cite{Cass}).

\section{$\pi + A$ Reactions}

Motivated by this observation we have performed calculations of the
dilepton invariant mass spectra in $\pi^-$ induced reactions on
nuclei \cite{Weidmann}; the experimental study of such reactions will be
possible in the near future at GSI. The calculations are based on a
semiclassical transport theory, the so-called Coupled Channel BUU
method (for details see \cite{Teis}) in which the nucleons, their
resonances up to 2 GeV mass and the relevant mesons are propagated from
the initial contact of projectile and target until the final stage of
the collision. This method allows one to
describe strongly-coupled, inclusive processes without any a-priori
assumption on the equilibrium or preequilibrium nature of the process.
Results of these calculations are being published in \cite{Weidmann}.

\begin{figure}
  \centerline{\includegraphics[height=9cm]{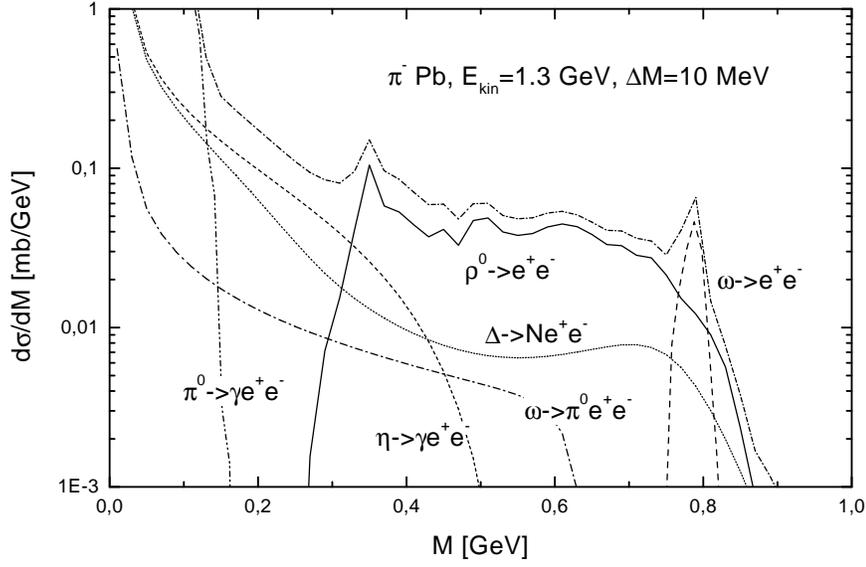}}
  \caption{Invariant mass yield of dileptons produced in pion-induced
  reactions at 1.3 GeV on Pb(from \protect\cite{Effe}). }
  \label{pidilept}
\end{figure}

Since then, we have improved the calculations in several aspects (see also
\cite{Knoll}). First,
we now include also processes like $N^* \rightarrow \Delta \rho$ in the
$\rho$ production channel. We find that this channel yields sizeable
contributions to the low-mass $\rho$ yield coming from the decay of rather
high-lying ($\approx 1.9$ GeV) nucleon resonances. Second, we now use
a more sophisticated treatment of broad resonances (i.e.\ the $\rho$), that
resembles that generally used for the treatment, e.g., of the $\Delta$
in transport calculations \cite{Effeabs}. Instead of producing the vector
mesons only at their pole mass like in \cite{Weidmann} and folding in the
phase-space weighted spectral function, we now
produce these particles already with the proper spectral function and
thus propagate also mesons on the tail of the mass-distribution. This
improvement, however, creates new problems: in these calculations very low
mass $\rho$ and $\omega$ mesons sometimes reach the nuclear surface and
escape, thus leading to a peak in the dilepton spectrum at very low masses.
(the structures in Fig.\ \ref{pidilept} below the vector meson mass are
due to statistical fluctuations and not to this effect).
We cure this problem at present only heuristically by including a
density-dependent selfenergy that ensures that all physical particles
freeze out with their vacuum spectral distribution. And finally, we now
use a VMD mandated $1/M^3$ weighting for the dilepton spectra, which is
consistent for calculations employing free hadron properties.

In these reactions the dominant dilepton emission channels are the same as
in ultrarelativistic heavy-ion collisions; this can be seen in Fig.
\ref{pidilept}
where I show the results for the dilepton spectra produced by
bombarding Pb nuclei with 1.3 GeV pions. Up to about 400 MeV
invariant mass the strongest component is given by the $\eta$ Dalitz
decay where the $\eta$'s are produced through the experimentally rather
well known process $\pi N \rightarrow \eta N$. Very close in magnitude
in this lower mass range is the somewhat model-dependent $\pi N$
bremsstrahlung component (not shown here). In the vector meson mass range
primarily both $\rho$ and $\omega$ contribute to the dilepton yield.
First, still preliminary calculations indicate that the in-medium
changes in the dilepton mass spectrum are of the order of a factor 2 in the
mass range between 300 MeV and 700 MeV \cite{Effe}.

There are various interesting problems in this process that ultimately
have to be answered by experiment. First, the $\pi N$ bremsstrahlung is
quite uncertain. Parts of it are included in a calculation that includes
processes like $\pi + N \rightarrow N^* \rightarrow N e^+e^-$ ($s$-channel
contributions), but $t$-channel processes for $\pi N$ scattering are not
so easy to handle, because the frequently used long-wavelength approximation
is known to be quite unreliable \cite{Lichard}. An exclusive measurement of
pion-induced
dilepton emission from the proton would be highly desirable to investigate
this point.

A further interesting problem is how to determine the branching ratios of
vector mesons into the dilepton channel. Simple VMD relates the coupling
strength to the bare (pole) mass $m_\rho$ of the meson so that one obtains
$\Gamma_{\rho \rightarrow e^+e^-}(M) \sim 1/M^3$. However, in
medium the vector meson is strongly coupled,
e.g.\, to $N^* h$ excitations, which may lead to mass-dependent vertex
corrections.

\section{Photonuclear Reactions}

Pion induced reactions have the disadvantage that the projectile
already experiences strong initial state interactions so that many
produced vector mesons are located in the surface where the
densities are low. A projectile that is free of this undesirable
behavior is the photon.

\subsection{Dilepton Production}
We have therefore -- also in view of a corresponding proposal for such
an experiment at CEBAF \cite{Preed} -- performed calculations for the
dilepton yield expected from $\gamma + A$ collisions.

\begin{figure}
  \centerline{\includegraphics[height=13cm]{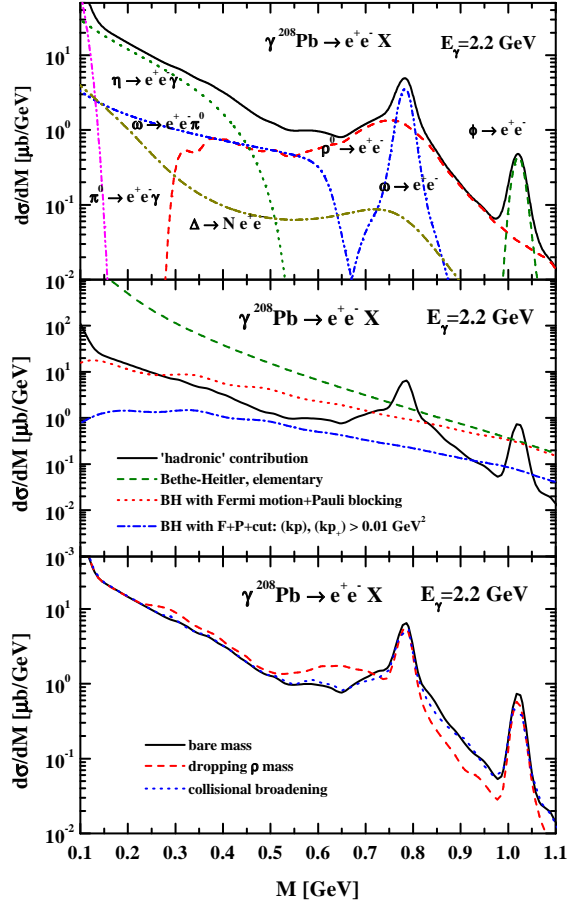}}
  \caption{Invariant mass spectra of dileptons produced in $\gamma +
  \mbox{}^{208}Pb$ reactions at 2.2 GeV. The top figures shows the
  various radiation sources, the middle figure the total yield from the
  top together with the Bethe-Heitler contributions, and the bottom part
  shows the expected in-medium effects (from \protect\cite{BratEffe}).}\label{gammadil}
\end{figure}

Results of these calculations are shown in Fig. \ref{gammadil}.
In the top figure the various sources of
dilepton radiation are shown. The dominant sources are again the same
as those in pion- and heavy-ion induced reactions, but the (uncertain)
$\pi N$ bremsstrahlung does not contribute in this reaction. The middle
part of this figure shows both the Bethe-Heitler (BH)
contribution and the contribution from all the hadronic sources. In the
lowest (dot-dashed) curve we have chosen a cut on the product of the
four-momenta of incoming photon ($k$) and lepton ($p$) in order to
surpress the BH contribution.
It is seen that even without BH subtraction the vector meson signal
surpasses that of the BH process.

The lowest
figure, finally, shows the expected in-medium effects \cite{BratEffe}: the
sensitivity in the region between about 300 and 700 MeV amounts to a
factor of about 3 and is thus in the same order of magnitude as in the
ultrarelativistic heavy-ion collisions.

In addition, the calculated strong differences in the in-medium
properties of longitudinal and transverse vector mesons can probably
only be verified in photon-induced reactions, where the incoming
polarization can be controlled; this is true also for dilepton
production with virtual, spacelike photons from inelastic electron
scattering. Another approach
would be to measure the coherent photoproduction of vector mesons; here
the first calculation available so far \cite{PE} shows a distinct
difference in the production cross sections of transverse and
longitudinal vector mesons.

\subsection{Photoabsorption}

Earlier in this paper I have discussed that a strong change of the
$\rho$ meson properties comes about because of its coupling to $N^* h$
excitations and that this coupling -- through a higher-order effect --
in particular leads to a very strong increase of the $\rho$ decay width
of the $N(1520) \: D_{13}$ resonance.

This increase may provide a reason for the observed disappearance of
the higher nucleon resonances in the photoabsorption cross sections on
nuclei \cite{Bianchi}. These cross sections scale very well with the
massnumber $A$ of the nucleus, which indicates a single-nucleon
phenomenon. They also clearly show a complete absence of the
resonance structure in the second and third resonance region. The
disappearance in the third region is easily explainable as an effect of
the Fermi-motion. The disapparance of the second resonance
region, i.e. in particular of the $N(1520)$ resonance, however, presents
an interesting problem; it is obviously a typical in-medium effect.

\begin{figure}
  \centerline{\includegraphics[height=7cm]{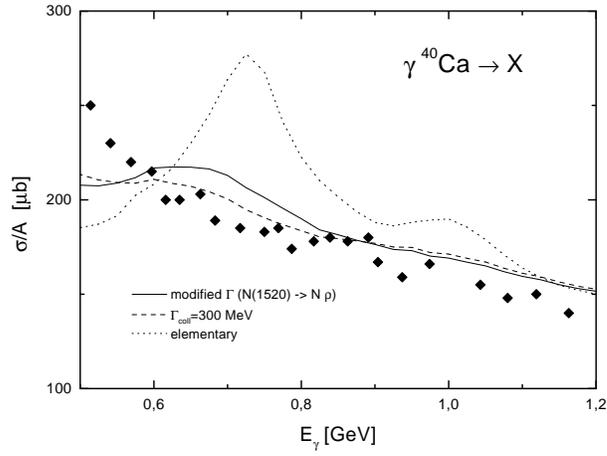}}
  \caption{Photoabsorption cross section in the second resonance region.
  Shown are the data from ref.\protect\cite{Bianchi}, the free
  absorption cross section on the proton, a Breit-Wigner fit with a total
  width of 300 MeV (dashed curve) and the result of a
  transporttheoretical calculation\protect\cite{Effeabs} with a medium
  broadened $\rho$ decay width of the $N(1520)$(solid curve).}\label{photabs}
\end{figure}

First explanations \cite{Kon} assumed a very strong collisional
broadening, but in ref. \cite{Effeabs} it has been shown that this
broadening is not strong enough to explain the observed disappearance
of the $D_{13}$ resonance. Since at the energy around 1500 MeV also the
$2 \pi$ channel opens it is natural to look for a possible connection
with the $\rho$ properties in medium. Fig.\ref{photabs}
shows the results of such an analysis (see also \cite{PE}). It is
clear that the opening of the phase space for $\rho$ decay of this
resonance provides enough broadening to explain its disappearance.

\section{Summary}
In this talk I have concentrated on a discussion of the in-medium
properties of the $\rho$ meson. I have shown that the scattering
amplitudes of the $\rho$ meson on nucleons determine the in-medium
spectral function of the $\rho$, at least in lowest order in the
nuclear density. The dilepton channel can give information on the
properties of the
$\rho$ deep inside nuclear matter whereas the $2 \pi$ decay channel --
because of its strong final state interaction -- can give only
information about the vector meson in the low-density region.

The original QCD sum rule predictions of a lowered
$\rho$ mass have turned out to be too naive, because they were based on
the assumption of a sharp resonance state. In contrast, all specific
hadronic models yield very broad spectral functions for the $\rho$
meson with a distinctly different behavior of longitudinal and transvese
$\rho$'s. Recent QCD sum rule analyses indeed do not predict a lowering
of the mass, but only yield --
rather wide -- constraints on the mass and width of the vector mesons.
I have also discussed that hadronic models that include the coupling of
the $\rho$ meson to nucleon resonances and a corresponding shift of
vector meson strength to lower masses give a natural backreaction on
the width of these resonances. In particular, the $N(1520) D_{13}$
resonance is significantly broadened because of its very large coupling
constant to the $\rho N$ channel. Since the $\rho$ decay
of this resonance has never been directly seen in an experiment, but is
deduced only from one partial wave analysis of $2 \pi$ production
\cite{Manley},
it would be very essential to have new, better data (and a new analysis of
these) for this channel.

A large part of the unexpected surplus of dileptons
produced in ultrarelativistic heavy-ion collisions can be understood by
simply including secondary interactions, in this case $\pi + \pi
\rightarrow \rho$, in the `cocktail plot' of hadronic sources; this
special source is one that is particular to heavy-ion collisions
with their very large pion multiplicity. Since the pions are produced
rather late during the collision, the gain from the high densities reached
in heavy ion collisions is not so large as one could naively have expected.

Motivated by this observation I have then discussed predictions for
experiments using pion and photon beams as incoming particles. In both
cases the in-medium sensitivity is nearly as large as it is in the
heavy-ion experiments. In addition, such experiments have the great
advantage that they take place much closer to equilibrium, an
assumption on which all predictions of in-medium properties are based.
Furthermore, only in such experiments it will be possible to look for
polarization effects in the in-medium properties of the $\rho$ meson.

I have finally shown that the in-medium properties of the $\rho$ also
show up in photonuclear reactions. One intriguing suggestion is that
the observed disappearance of the second resonance region in the
photoabsorption cross section is due to the broadening of the $N(1520)$
resonance caused by the shift of $\rho$ strength to lower masses.

At high energies, finally, the in-medium broadening of the $\rho$ leads
to a mean free path of about 2 fm in nuclear matter. Shadowing will
thus only be essential if the coherence length is larger than this mean
free path. Increasing the four-momentum transfer $Q^2$ at fixed energy
transfer $\nu$ leads to smaller coherence length and thus diminishes
the initial state interactions of the photon leading to a larger
transparancy. This effect is essential to verify experimentally; it is
superimposed on the color-transparency effect that is still being
looked for.

\section{Acknowledgement}
This talk is based on results of work with E. Bratkovskaya, W. Cassing, M.
Effenberger, H. Lenske, S. Leupold, W. Peters, M. Post and T. Weidmann. I
am grateful to all of them for many stimulating discussions. I wish to
thank in particular M. Effenberger and E. Bratkovskaya for providing me
with the results shown in Figs. 4 and 5 before publication.





\end{document}